\documentclass{article}

\usepackage{arxiv}

\usepackage[utf8]{inputenc} 
\usepackage[T1]{fontenc}    
\usepackage{hyperref}       
\usepackage{url}            
\usepackage{booktabs}       
\usepackage{amsfonts}       
\usepackage{nicefrac}       
\usepackage{microtype}      
\usepackage{lipsum}		
\usepackage{graphicx}
\usepackage[square,numbers]{natbib}
\usepackage{doi}

\title{Social media sharing by political elites: An asymmetric American exceptionalism}

\date{} 					

\author{ \href{https://orcid.org/0000-0002-4274-4580}{\includegraphics[scale=0.06]{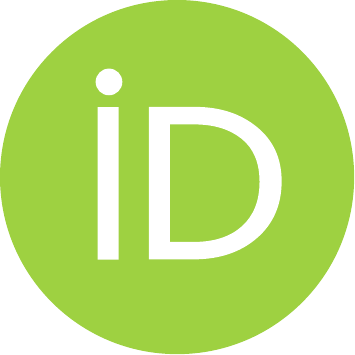}\hspace{1mm}Jana Lasser} \\
	Graz University of Technology\\
	Complexity Science Hub Vienna\\
	\And
	\href{https://orcid.org/0000-0002-2571-3731}{\includegraphics[scale=0.06]{orcid.pdf}\hspace{1mm}Segun Taofeek Aroyehun} \\
	Graz University of Technology\\
	\And
	\href{https://orcid.org/0000-0003-2629-2913}{\includegraphics[scale=0.06]{orcid.pdf}\hspace{1mm}Almog Simchon} \\
	University of Bristol\\
	\And
	\href{https://orcid.org/0000-0003-4918-3875}{\includegraphics[scale=0.06]{orcid.pdf}\hspace{1mm}Fabio Carrella} \\
	University of Bristol\\
	\And
	\href{https://orcid.org/0000-0002-2820-9151}{\includegraphics[scale=0.06]{orcid.pdf}\hspace{1mm}David Garcia} \\
	Graz University of Technology\\
	Complexity Science Hub Vienna\\
	\And
	\href{https://orcid.org/0000-0003-1655-2013}{\includegraphics[scale=0.06]{orcid.pdf}\hspace{1mm}Stephan Lewandowsky} \\
	University of Bristol\\
	University of Western Australia\\
}

\renewcommand{\headeright}{Preprint}
\renewcommand{\undertitle}{Preprint}
\renewcommand{\shorttitle}{Social media sharing by political elites: An asymmetric American exceptionalism}

\hypersetup{
pdftitle={Social media sharing by political elites: An asymmetric American exceptionalism},
pdfsubject={CS.CY},
pdfauthor={Jana Lasser},
pdfkeywords={misinformation, elites, political discourse},
}

\begin{document}
\maketitle

\begin{abstract}
	Increased sharing of untrustworthy information on social media platforms is one of the main challenges of our modern information society. Because information disseminated by political elites is known to shape citizen and media discourse, it is particularly important to examine the quality of information shared by politicians. Here we show that from 2016 onward, members of the Republican party in the U.S. Congress have been increasingly sharing links to untrustworthy sources. The proportion of untrustworthy information posted by Republicans versus Democrats is diverging at an accelerating rate, and this divergence has worsened since president Biden was elected. This divergence between parties seems to be unique to the U.S. as it cannot be observed in other western democracies such as Germany and the United Kingdom, where left-right disparities are smaller and have remained largely constant.
\end{abstract}


\section*{Introduction}
Elite cues are important drivers of public public opinions and discourse~\cite{chong2007theory}. 
For example, public concern about climate change is largely influenced by opinions and information shared by elites~\cite{carmichael2017elite}.
The public's increased polarisation over climate change reflects the retreat of the Republican leadership
from the scientific evidence~\cite{merkley2018party}. 
The power of elites to set the agenda of public conversations extends to mainstream media. For example, 
Donald Trump has been shown to successfully divert media attention 
away from topics that were potentially harmful to him~\cite{lewandowsky2020using}. 
Notwithstanding the importance of elite discourse, 
research attention has only recently shifted to investigating the information sharing practices of a 
broader range of members of the political elite~\cite{mosleh2021falsehood}. 
Here we contribute to these analyses by investigating the quality of information shared on social media by members of the U.S. Congress. 
We find that the trustworthiness of information shared by members of the Republican party is declining and that this decline has accelerated after the election of Joe Biden as president. We contrast these findings with 
information sharing practices of political elites in two other western democracies, 
Germany and the United Kingdom. In both cases, although parties on the political right tend to share somewhat
less trustworthy information, the rapid decline of information quality shared by Republicans during the last 6 years 
is unique to the U.S. and is not observed in other countries.

\section*{Results}
To assess the trustworthiness of information shared by politicians with the general public, we retrieve three corpora of tweets: by former and active members of the U.S. Congress, the German parliament and the British parliament. For each corpus, we retrieve all tweets posted between January 1, 2016 and March 16 2022. We do not include retweets. We extract all URLs included in the tweets. We follow an approach employed by similar research in this domain~\cite{grinberg2019fake, pennycook2021shifting} and use a trustworthiness assessment by professional fact checkers of the \emph{domain} a link points to. To this end, we use the NewsGuard database~\cite{newsguard2022}. As of March 2022, NewsGuard indexes 6860 English and 145 German language domains. 
Each domain is scored on a scale of 0 (very poor) to 100 (exceptional quality journalism) points. 
Domains with less than 60 points are considered ``not trustworthy''~\cite{newsguard2022}. 
The majority of indexed domains (62.8\% for English, 74.5\% for German) are considered trustworthy. 
After excluding links to social media websites (Twitter, YouTube, Facebook, Instagram) and search engines (Google, Yahoo), 
the database covers 50.5\% of links posted by members of the U.S. Congress, 54.6\% of links posted by members of the German parliament and 44.2\% of links posted by members of the British parliament. 
Coverage generally increases slightly over time and is similar between parties (see Materials and Methods for details). 

\begin{figure*}[!ht]
    \centering
    \includegraphics[width=\textwidth]{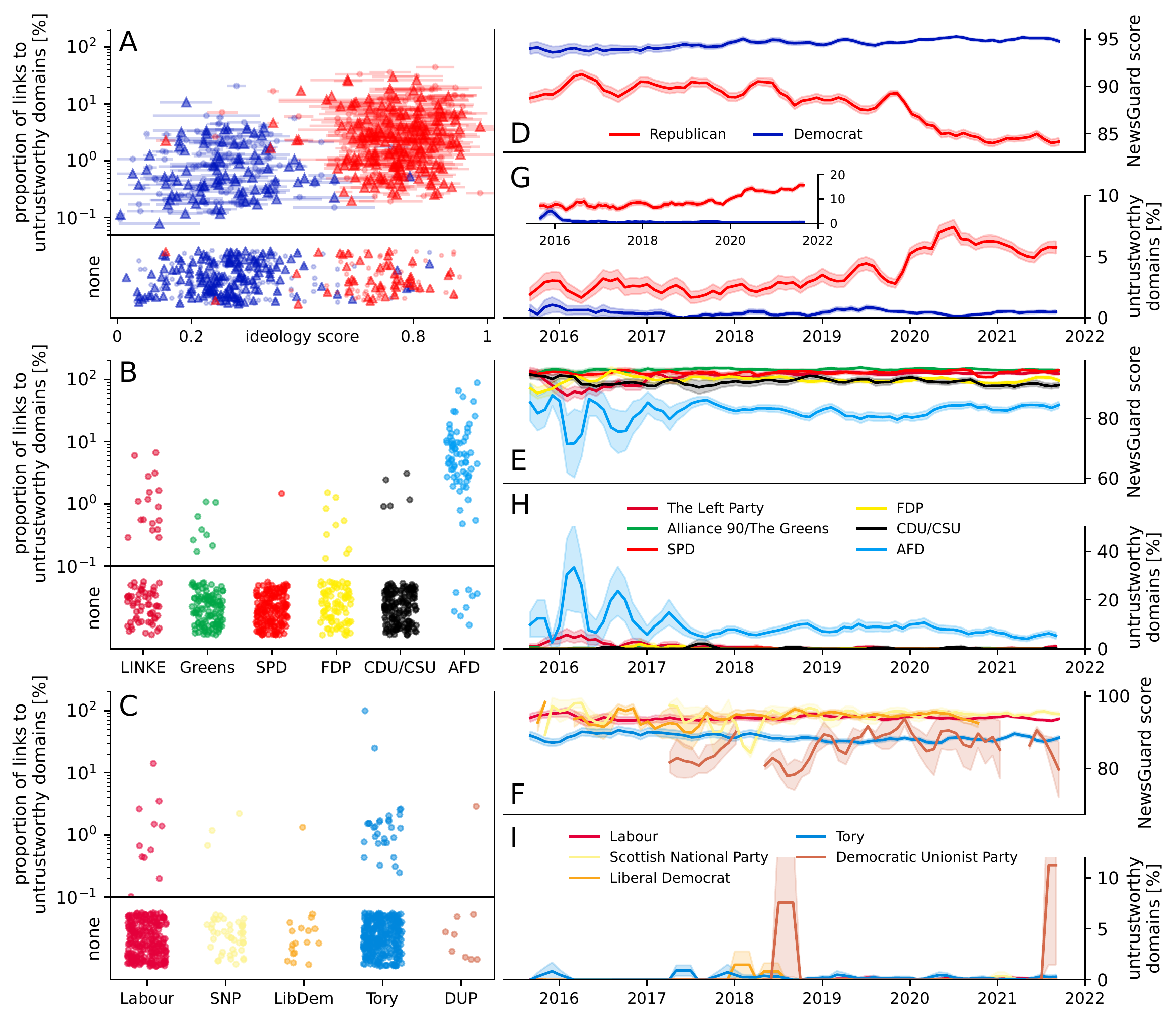}
    \caption{Proportion of links to untrustworthy domains posted by Twitter accounts associated with Democratic and Republican former (dots) and active (triangles) members of the U.S. Congress (\textbf{A}), members of the German (\textbf{B}) and British parliament (\textbf{C}). Average NewsGuard score of links posted by members of the U.S. Congress (\textbf{D}), members of the German parliament (\textbf{E}) and members of the British parliament (\textbf{F}) between 2016 and 2022. Proportion of links to untrustworthy domains posted by members of the U.S. Congress (\textbf{G}), members of the German parliament (\textbf{H}) and members of the British parliament (\textbf{I}) between 2016 and 2022. The inset in \textbf{G} shows a reproduction of the result using an independently compiled list of untrustworthy domains (see materials and methods). Scores and proportions of untrustworthy domains were averaged over monthly intervals with a rolling average of three months, and are broken down by party, colour-coded by commonly used party colours. The 95\% confidence intervals were computed with bootstrap sampling over 1,000 iterations.}
    \label{fig:fig1}
\end{figure*}

We report the overall proportion of links that point to domains that are considered untrustworthy, as well as the NewsGuard score. Figure~\ref{fig:fig1} \textbf{A}-\textbf{C} shows the proportion of links to untrustworthy domains for the three countries. For the U.S., we report values over ideology scores provided by GovTrack~\cite{govtrack2022}, for Germany and the United Kingdom we report values broken down by parties. Republicans share more untrustworthy information than Democrats (note the logarithmic scale). For Germany, parties on the extreme left and extreme right share more untrustworthy information than parties in the centre. Overall, Republicans share 9.1 times more links to websites considered unstrustworthy than Democrats (Republicans 3.87\%, Democrats 0.42\%, difference 3.45\%, 1.83 SD). For Germany, members of the CDU/CSU post 6.1 times more links to such websites than members of the SPD (CDU/CSU 0.19\%, SPD 0.03\%, difference 0.16\%, 0.02 SD). For the UK, members of the Tories post 4.7 times more links to untrustworhty domains than members of the Labour party (Tory 0.23\%, Labour 0.05\%, difference 0.18\%, 0.07 SD). For both Germany and the UK, the conservative parties post more links to untrustworthy domains than their counterparts on the left, but overall they post about half as many such links as the Democrats in the U.S. We also note that numbers for Germany and the UK are based on very low overall counts of links to untrustworthy domains.

In Fig.~\ref{fig:fig1} \textbf{D}-\textbf{F} we show the temporal trend of the NewsGuard score, 
averaged over all links posted in a given month broken down by party. Links posted by Republicans show a notable decrease in trustworthiness, from on average $89.9\pm 0.1$ (mean $\pm$ SD) points in the years 2016-2018 to $85.2\pm 1.7$ points in the years 2020-2022. The score of links posted by Democrats stays remarkably stable ($94.2 \pm 0.4$ in 2016-2018 and $94.9\pm 0.1$ in 2020-2022). This development is not reflected in the trustworthiness scores of links posted by conservative or far-right parties in other countries. In Germany, the scores of links posted by the AfD and members of the CDU/CSU stays stable with an average score of $83.4\pm 1.0$ and $92.2\pm 0.7$ in 2016-2018 and a score of $83.2\pm 1.4$ and $91.6 \pm 0.9$ in 2020-2022, respectively. Similarly, the scores of the Democratic Unionist Party and Conservatives in the British parliament stay stable or slightly improve, with $83.4 \pm 1.0$ and $89.1\pm 0.6$ in 2016-2018 and $85.6\pm 5.4$ and $88.2\pm 0.3$ in 2020-2022, respectively. These overall trends are also reflected in the proportion of links to domains 
considered untrustworthy (score $< 60$), shown in Fig.~\ref{fig:fig1} \textbf{G}-\textbf{I}. The proportion of links to these domains posted by Republicans doubles, from $2.4\pm 0.2$\% in 2016-2018 to $5.5\pm 0.6$\% in 2020-2022. The proportion of untrustworthy links posted by Democrats shows no change, from $0.4\pm 0.3$\% in 2016-2018 to $0.4\pm 0.1$\% in 2020-2022. For the German AfD, the proportion decreases from $8.8\pm 2.2$\% to $6.8\pm 2.0$\%. The proportion of links to untrustworthy domains of all other parties in Germany and the United Kingdom is similar to the proportion posted by Democrats and does not change over time. \\
As a robustness check, we reproduced our main result (viz. the increase in the proportion of links to untrustworthy domains by
Republicans) using a second database of domain trustworthiness, compiled independently of NewsGuard (see Materials and Methods for details). The observed temporal trend of the proportion of links to untrustworthy domains is displayed in the inset of Fig.~\ref{fig:fig1} \textbf{G}, and shows a similar trend in the proportion of untrustworthy domains posted by Republicans (from $5.8\pm 0.3$\% in 2016-2018 to $11.3\pm 2.7$\% in 2020-2022), while the proportion of links to untrustworthy domains posted by Democrats slightly decreases ($1.0\pm 0.9$\% and $0.4\pm 0.1$\%).

\section*{Discussion}
Several recent analyses have shown that American conservatives are more likely to encounter and share untrustworthy information than their counterparts on the political left~\cite{grinberg2019fake,Guess20a,Guess19}. Although the reasons for this apparent asymmetry are still
debated, one possible explanation appeals to partisan motivations. Evidence suggests that derogatory content towards the political outgroup increases sharing intentions among partisans ~\cite{Rathje2021-fn}. According to a recent study, greater negativity towards Democrats is mostly found in lower-quality outlets, which may explain conservatives’ over-representation in sharing untrustworthy information ~\cite{Osmundsen2021-gn}. Another possibility is that right-wing actors leverage controversial outlets in order to get more traction on social media, as has been evident in the US, UK, and Germany~\cite{Huszar2022-aa}.

Here we contribute to a potential explanation by showing that Republican members of Congress have become increasingly likely to share untrustworthy information on Twitter. This pattern can contribute to the observed asymmetry among the public in at least two ways: first, by directly providing misinformation to Republican partisans and, second, by legitimizing the sharing of untrustworthy information more generally.  Notably, this pattern is less pronounced in two major western democracies: although politicians from mainstream conservative parties in both Germany and the UK tend to share information that is of slightly lower quality than information shared by their counterparts on the left, the gap is not large and there is no evidence of it widening. We are thus experiencing a uniquely American dilemma.

\section*{Methods}
\subsection*{Twitter corpus}
A corpus of tweets from former and present
members of the U.S. Congress, the German parliament and the British parliament was collected by scraping all 
tweets from accounts associated with the respective politicians between January 1, 2016 and March 16, 2022. Lists of Twitter handles of the 114$^\mathrm{th}$ to 117$^\mathrm{th}$ Congress were collected from a number of sources\footnote{\url{https://www.socialseer.com/resources/us-senator-twitter-accounts/}}\footnote{\url{https://dataverse.harvard.edu/dataset.xhtml?persistentId=doi:10.7910/DVN/MBOJNS}}\footnote{\url{https://triagecancer.org/congressional-social-media}}. For the 114$^\mathrm{th}$ and 115$^\mathrm{th}$ Congress, only handles of senators were available. This resulted in a total of 1,143 unique Twitter handles, which includes congressional staff and congress member campaigns. 108 accounts were not accessible because they had been deleted, suspended, or set to ``private''. Lists of Twitter handles for members of the German and British parliament were collected from two online sources\footnote{\url{https://twitter.com/i/lists/912241909002833921} | \url{https://www.politics-social.com/list/followers}} and the Twitter Parliamentarian Database~\cite{van2020twitter}, resulting in a total of 823 and 727 unique Twitter handles, respectively. To build the text corpus, all tweets posted by the collected Twitter accounts in the specified time frame were obtained via
the Twitter API. We chose January 1, 2016 as the earliest date because before this date very few tweets by parliamentarians in Germany and the United Kingdom were available. 
This resulted in a total of 1,696,626, 754,233 and 960,114 tweets for the US, Germany, and U.K., respectively. 
To determine the trustworthiness of information posted by the politicians, we extracted all URLs linking external sites contained in the tweets (shortened links such as bit.ly were expanded to determine the actual domain).

\subsection*{NewsGuard scores}
Following the methods of prominent research concerned with the trustworthiness of information~\cite{grinberg2019fake, pennycook2021shifting}, we use source trustworthiness as an estimator for the trustworthiness of an individual 
piece of shared information. Specifically, we classified the trustworthiness of links tweeted by politicians based on the trustworthiness of the domain rather than specific items of content. We used nutrition scores provided by NewsGuard, a company that offers professional fact checking as a service and curates a large data base of domains. The trustworthiness of a domain is assessed on a scale of 0 to 100 points, where domains with a score of 60 or higher are labelled as ``generally adhering to basic standards of credibility and transparency''~\cite{newsguard2022}. 
Similar to~\cite{bhadani2022political}, we use this value as a threshold below which we classify a domain as ``not trustworthy''. 
 
\subsection*{Independent data base of domain trustworthiness}
To validate our use of NewsGuard we compiled an independent data base of domain trustworthiness from a range of academic and journalistic fact checking sites. Most of these sources were also used by~\cite{gallotti2020assessing}. The list includes Bufale~\cite{bufale}, Bufalopedia~\cite{bufalopedia}, Butac~\cite{butac}, Buzzfeed News~\cite{buzzfeednews}, Columbia Journalism Review~\cite{columbiajournalism}, Fake News Watch~\cite{fakenewswatch},
Media Bias Fact Check~\cite{mediabiasfactcheck}, Politifact~\cite{politifact}, and Melissa Zimdars~\cite{zimdars}. After merging these lists and removing duplicates, the combined list contains 4767 domains. A total of 1677 of these domains are also contained in the NewsGuard data base, as of March 1, 2022. 

The main challenge in combining lists from different fact checkers lies in unifying the labels the fact checkers assign to the domains. To address this, we devise a scheme where we rate each domain on two dimension that we consider to be important to assess reliability and trustworthiness of information: "accuracy" and "transparency". We devise an accuracy scale that varies from 1 (false information) to 5 (scientific) and a transparency scale that varies from 1 (no transparency) to 3 (transparent). We provide a more detailed description of the five accuracy levels as well as mappings of the labels of individual fact checking sites to accuracy and transparency scores and the full list of domains at \url{https://doi.org/10.5281/zenodo.6536692}.

After mapping all individual lists to the "accuracy" and "transparency" dimensions, we label every domain that has an accuracy score of 1 (False Information) or 2 (Clickbait) and/or a transparency score of 1 (No Transparency) as "unreliable". This results in a total of 2170 domains being labelled as "unreliable" and 2597 as "reliable". For the 1677 domains that are contained in both data bases, the Krippendorff's $\alpha$ between "untrustworthy" (score $< 60$ in NewsGuard) and "unreliable" in the independently compiled data base is 0.84, which shows a very high agreement between the two data bases. 
The independent list contains 4,767 unique domains, of which 2,170 are labelled as ``untrustworthy''. Less then half of the domains ($N=1,677$, of which 615 are untrustworthy) are also contained in the NewsGuard data base. We publish the independently compiled domain list for other researchers to use\footnote{https://doi.org/10.5281/zenodo.6536692}.

\section*{Acknowledgements}
JL acknowledges funding by the Marie Skłodowska-Curie grant No. 101026507. SL, AS, STA and DG received support from the European Research Council (ERC Advanced Grant 101020961 PRODEMINFO). SL and AS acknowledge support from the Volkswagen Foundation (grant ``Reclaiming individual autonomy and democratic discourse online: How to rebalance human and algorithmic decision making''), and SL, DG, and FC also acknowledge funding from the John Templeton Foundation (through the Honesty program awarded to Wake Forest University). SL was also supported by the Humboldt Foundation through a research award.

\section*{Data Archival}
Full reproduction materials including data (tweet IDs, party labels, independently compiled list of domain labels) and analysis code but excluding the NewsGuard data base which is proprietary
are accessible at \url{https://doi.org/10.17605/OSF.IO/MQHGP}.

\bibliographystyle{abbrvnat}
\bibliography{references,mega}
\end{document}